# Guided Probabilistic Simulation of Complex Systems Toward Rare and Extreme Events

Tarannom Parhizkar

*B. John Garrick Institute for the Risk Sciences, University of California, Los Angeles, 404 Westwood Plaza, Los Angeles, CA 90095, USA. E-mail: tparhizkar@ucla.edu*

Ali Mosleh

*B. John Garrick Institute for the Risk Sciences, University of California, Los Angeles, 404 Westwood Plaza, Los Angeles, CA 90095, USA. E-mail: Mosleh@ucla.edu*

*SUMMARY*

Simulation-based or dynamic probabilistic risk assessment methodologies were primarily developed for proving a more realistic and complete representation of complex systems accident response. Such simulation-based methodologies have proven to be particularly powerful for systems with control loops and complex interactions between its elements, be they hardware, software, or human, as they provide a natural probabilistic environment to include physical models of system behavior (e.g., coupled neutronics and thermal-hydraulic codes for nuclear power plants), mechanistic models of materials or hardware systems to predict failure, and those of natural hazards. Despite the advancements in simulation-based methodologies, the fundamental challenge still persists as the space of possible probabilistic system trajectories is nearly infinite in size in simulating even systems of relatively low complexity.

Existing methodologies that analyze and predict these scenarios, some of which may involve rare events of interest, are either computationally prohibitive for the high-dimensional systems of interest or are only effective for low-dimensional dynamics and cannot capture the large-scale, nonlinear behavior of interconnected complex systems.

In this paper, a framework is developed to identify rare and extreme events and enabling the use of reverse trajectories to trace failures (or other system states) to causes for potential mitigation actions. This framework consists of an Intelligent Guidance module, Trajectory Generation module and Physical Simulation module. The Intelligent Guidance module provides planning information to the Trajectory Generation module that creates scenarios by interacting with the Physical Simulation in its environment. In turn, system trajectories or scenarios are created and post-processed to provide updating information to the Intelligent Guidance module or aggregate the results when stopping criteria are met. The objective of guided simulation is to control the growth of the "scenario tree" and to efficiently identify important scenarios that meet single or multiple criteria. We present several solution strategies, both qualitative and data-driven for each module.

## 1 INTRODUCTION

Complex systems, whether natural or manmade, pose formidable challenges in predictability. These challenges have two main dimensions: complexity itself in terms of topological, functional, and behavioral features, and limitation in data and knowledge needed to understand their behavior. The system behaviors are often not fully understood. By definition, our understanding captured through models of the real world is limited. Therefore, our models of reality are subject to knowledge-driven or epistemic uncertainty. Real world phenomena also exhibit variability inherent to the natural phenomena, so our models are also subject to stochastic or aleatory uncertainty.

In many applications, the primary interest is to find trajectories leading to a specific state or set of the states of the system. In certain applications the focus is on identifying the trajectories that lead to extremely rare but highly significant system states (e.g., unanticipated catastrophic failures). This situation poses a formidable challenge as such trajectories are often at or outside the boundaries of the scientific and engineering knowledge, and also could easily be masked by model abstractions and solution techniques. It is not difficult to note the limitations of purely data-driven approaches such as statistical techniques (e.g., extreme value theory) or even more modern data analytics when predicting and identifying trajectories to rare system states that are of interest. The solution typically requires a model-based (e.g., causal models) approach, where model parameter assessment and model validation would be informed by data, or more generally all the available evidence.

Important to the assessment of rare and extreme events are the limitations of purely data-driven approaches and the benefits of model-based approaches using dynamical systems theory and probabilistic techniques to capture stochastic phenomena possibly augmented by latest generation computational techniques such as adaptive sampling and reinforcement learning. The simulation-based methodologies offer several advantages over the conventional approaches currently used by many industries worldwide. These advantages include time-dependent prediction and quantification of the operator error-forcing contexts, better representation of the deterministic success criteria, and considerable reduction in analyst-to-analyst variability of the results.

Simulation-based or dynamic probabilistic risk assessment methodologies were primarily developed for proving a more realistic and complete representation of the complex systems accident response because, as history has shown multiple times already, the contextual details are important. In practice, these details may not be covered by the regulations or standards as systems, even of similar design, could be unique in their own way with different operators, site-specific components, and changing environmental conditions in which they are deployed. For example, in the case of complex industrial plants, of importance to the safety of the public and environment are the off-site hazards in terms of immediate or latent health effects and land contamination.

Simulation-based probabilistic risk assessment methodologies are model-based simulations that generate system trajectories (i.e., risk scenarios) with their associated probabilities of occurrence. To achieve this, rules of deterministic and stochastic behaviors of the complex system and its elements, such as hardware, software, human operators, process variables, and environmental conditions, are developed and implemented as building blocks of a computer simulation framework. Each simulation module tracks all the possible changes in the functional states and parameters associated with each element or the whole system as a function of time. By accounting for the nature and impact of the interactions and interdependencies between elements at any level of the system, risk scenarios are generated by a simulation engine. Depending on the particular method chosen for scenario generation, the probabilities of each scenario or clusters of scenarios are calculated for the system end states of interest.

Such simulation-based methodologies have proven to be particularly powerful for systems with control loops and complex interactions between its elements, be they hardware, software, or human, as they provide a natural probabilistic environment to include physical models of system behavior (e.g., coupled neutronics and thermal-hydraulic codes for nuclear power plants), mechanistic models of materials or hardware systems to predict failure, and those of natural hazards. Most of the simulation-based probabilistic risk assessment methodologies fall under two main categories: continuous-time, such as Continuous Event Tree (CET) [1] [2] and discrete-time, such as Dynamic Event Tree Analysis Method (DETAM) [3] [4], Accident Dynamic Simulator (ADS) [5] [6], Analysis of Dynamic Accident Progression Trees (ADAPT) [7], and Risk Analysis and Virtual Environment (RAVEN) [8]. The continuous-time methodologies simulate the possible dependencies between the process variables, hardware, software, and human interactions with a single integral equation, generally solved with Monte Carlo techniques. The discrete-time methodologies use discrete dynamic event trees (DDETs) that are computationally generated based on time-dependent models of system evolution and various branching conditions. Essentially, all discrete methodologies employ a simulation engine that generates branches at certain time steps, known or unknown a priori, with their associated probabilities and computes the probability of each scenario. Branching points can include system hardware states, physical variable changes, human actions, software failures, or an end state if one of the stopping criteria is met. A graphical representation of the system evolution space with its probabilities of occurrence is shown in Figure 2.

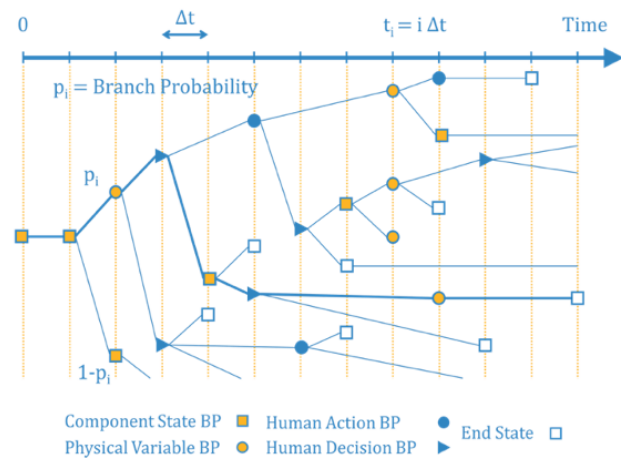

*Figure 2. Discrete dynamic event tree*

A very large effort is underway to develop a comprehensive simulation environment for risk analysis called Risk Analysis and Virtual Environment (RAVEN). It is part of the Multiphysics Object Oriented Simulation Environment (MOOSE) [9], a parallel finite element framework for integrating fully coupled, fully-implicit multi-physics solvers. The fundamental idea behind RAVEN is to employ various sampling strategies to perturb the timing and sequencing of events, initial conditions, and internal parameters of physical models to estimate the probability of occurrence of unintended events with high consequence. Applied to small-scale problems, this approach is very powerful. Nevertheless, for complex systems the set of uncertain parameters that characterize these systems grows exponentially large. Therefore, even with the fairly large computational resources that are accessible nowadays, exploring the hyperspace with a good level of confidence is generally not affordable. Despite these advances, the fundamental challenge persists as the space of possible probabilistic system trajectories is nearly infinite in size in simulating even systems of relatively low complexity.

One approach is to guide the simulation toward rare and extreme events. The idea of leveraging high knowledge and heuristics level for guiding a complex system simulation was first introduced in the SimPRA platform developed for NASA [10] [11] [12]. In SimPRA, a hierarchical planning engine generates high level risk scenarios automatically. Entropy-based biasing techniques are used to adaptively guide the simulation towards events and end-states defined in the plan. Multi-level scheduling is used to adjust the levels of detail of the simulation elements. Prior knowledge of the system and knowledge gained during simulation are used to dynamically adjust the exploration rules. The simulation model used in SimPRA is a multi-level model. The behavior of system components is abstracted at different levels. The simulation's level of detail is adjusted automatically during the simulation based on heuristics and high-level principles. SimPRA was used to find possible anomalies of Space Shuttle flights during ascent to orbit and was shown to converge to states of interest much faster than simulations without planner features. An innovative aspect of the approach is that the selection of the most appropriate level of detail is initially specified by the analyst in the planner but is then automatically adjusted during the various rounds of simulation according to the entropy-based rule.

The primary objective of this paper is to develop a framework for identifying rare and extreme events and enabling the use of reverse trajectories to trace failures (or other system states) to causes for potential mitigation actions. This framework consists of three main modules including an Intelligent Guidance module, Trajectory Generation module and System Simulation module. In this paper, possible approaches that can be implemented in each module are presented, and the interaction between modules is discussed.

## 2 METHODOLOGY OVERVIEW

System evolution scenarios lead to, or go through, "system states" some of which may be characterized according to a set of defined performance criteria, for instance failure or success in meeting the functional objectives of the system. Typically, only a small subset of this space includes scenarios that lead to very rare and undesirable states. Recognizing this fact offers a potential pathway for reducing the size of the space of exploration. In the proposed framework, effective methods are proposed to only explore the part of the space of scenarios with a high potential for generating events that meet a predefined set of criteria. At first, we should define a unifying mathematical formalism such as one used in the Theory of Probabilistic Dynamics [1], captured by the following equation:

$$\frac{d\bar{r}}{dt} = f_{\bar{x}}(\bar{r})$$

$$\frac{\partial \pi(\bar{x},\bar{r},t|\bar{x}_0,\bar{r}_0,t_0)}{\partial t} + \nabla_{\bar{r}}[f_{\bar{x}}(\bar{r},t)\pi(\bar{x},\bar{r},t|\bar{x}_0,\bar{r}_0,t_0)]$$
$$+ \lambda_{\bar{x}}(\bar{r})\pi(\bar{x},\bar{r},t|\bar{x}_0,\bar{r}_0,t_0)$$
$$- \sum_{\bar{y} \neq \bar{x}} p(\bar{y} \to \bar{x}|\bar{r})\pi(\bar{x},\bar{r},t|\bar{x}_0,\bar{r}_0,t_0) = 0$$

This set of partial differential equations is a variation of the

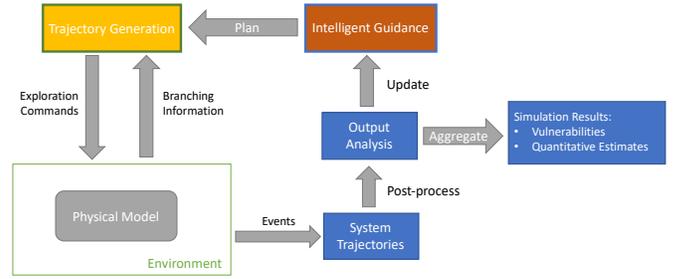

Chapman-Kolmogorov equation [13]. It is a first-order equation because it assumes a deterministic behavior between component transitions. This equation embodies the mathematics of continuous event trees (CET), or a set of system evolution trajectories. This, and similar formalisms, of course only provide a conceptual integrator of the various models and parameters governing the dynamic evolution of systems.

In this study, a conceptual framework is proposed that guides the integration for characterizing patterns of complex system behavior. The architecture of the guided simulation for capturing complex system behavior and guiding the simulation toward regions of high risk-significance is shown in Figure 3.

*Figure 1. Guided Simulation Architecture*

The framework includes Intelligent Guidance, Trajectory Generation, and Physical Simulation modules. The Intelligent Guidance module provides planning information to the Trajectory Generation module that creates scenarios by interacting with the Physical Simulation in its environment. In turn, system trajectories or scenarios are created and post-processed to provide updating information to the Intelligent Guidance module or aggregate the results when stopping criteria are met. These three modules are explained in the following sections.

## 3 INTELLIGENT GUIDANCE

The objective of guided simulation is to control the growth of the "scenario tree" and to efficiently identify important scenarios that meet single or multiple criteria. This module can include several solution strategies, both qualitative and data driven as presented in **Error! Reference source not found.**.

### 3.1 Guided Simulation via Qualitative Biasing

This approach uses an initial high-level map of possible scenarios (planner) as a qualitative seed for the more detailed Trajectory Generation algorithm.

### 3.2 Reinforcement Learning (RL)

Reinforcement learning is a field of Artificial Intelligence (AI) emerged to solve continuous and discrete sequential decision-making problems. We introduce phenomena from RL to guide simulations toward trajectories that have a high probability of reaching the undesirable states. We borrow the concept of state value function from RL to estimate the worthiness of each point in the space of simulation control variables and intrinsic system variables. The value of each trajectory is realized at the end of the trajectory, and then this value is backpropagated to all the points visited by the system throughout the trajectory by means of temporal difference learning which is the core idea of the most recent RL techniques [14].

### 3.3 Deep Reinforcement Learning (DRL)

DRL techniques can be used to deal with the high dimension of variables in large-scale, interconnected systems. Deep learning is known for its revolutionary capability of extracting patterns from high dimension, nonlinear data. Over the last decade, great progress has been achieved in utilizing deep learning in RL which results in highly intelligent DRL agents that could achieve superhuman performance in many simulated environments, namely video and Atari games and simulated self-driving cars. In this module, we can employ the state-of-the-art DRL techniques into complex system simulation to develop strategies that find the scenarios with the highest probability of reaching states of interest [15] [16] [17] [18].

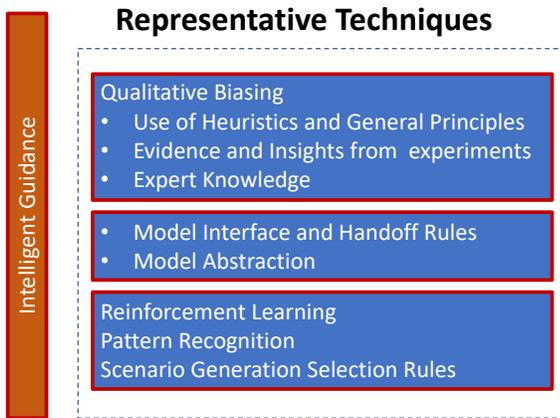

*Figure 4. Representative technique to be integrated into the Intelligent Guidance module*

## 4 TRAJECTORY GENERATION

Similar to the Intelligent Guidance, the Trajectory Generation module (Figure 1) controls the space explosion phenomenon and interacts with the Physical Simulation using a combination of possible solution strategies (**Error! Reference source not found.**).

### 4.1 Entropy-based Look Ahead

A few research efforts have shown that information entropy [19] can be used as an effective way to guide simulation-based prediction of complex system evolution and significantly reduce the needed computational resources or achieve a more complete solution space [20] [21] [22]. In essence, we plan to use information or diagnostic entropy to "look ahead" based on expected gain in information (e.g., possibility of reaching an undesirable state in the next few simulation steps) to then decide whether to go further, and which system state transition branches to explore. Entropy-based biasing technique offers a way to adaptively guide the simulation toward events and end-states of interest. This promising application of information entropy is still in early stages of development, however the recent advances in simulation capabilities should provide the right background in which such theory can mature in concert with the other techniques.

### 4.2 Exploration of Risk-Significant Regions

The exploration of risk-significant regions of the parametric input space with respect to various figures of merit, such as failure, has shown great promise for system responses from surrogate models. The need for surrogate models lies in the computationally prohibitive needs of such algorithms to reach convergence in large parametric domains. One of the limitations of this approach is the very challenging validation effort required due to the inherent nature of surrogate models. In this regard, model-based adaptive sampling strategies can be used to fill the gap between purely data-driven and model-based approaches.

### 4.3 Scaling-effects on uncertainty

Phenomenologically accurate and computationally fast and efficient simulations over a broad range of scenarios and scales are a must. Advanced mechanistic first-principal models and simulations provide higher accuracy than ever enabled by high-performance computing and data storage, and advanced measurement and diagnostics. Leaving aside the challenge of coupling and scaling up the models, the impact of uncertainty due to scaling needs to be assessed as well. In this module, the impact of uncertainty due to scaling from high-resolution nano- or micro-scale to the system-level scale can be studied.

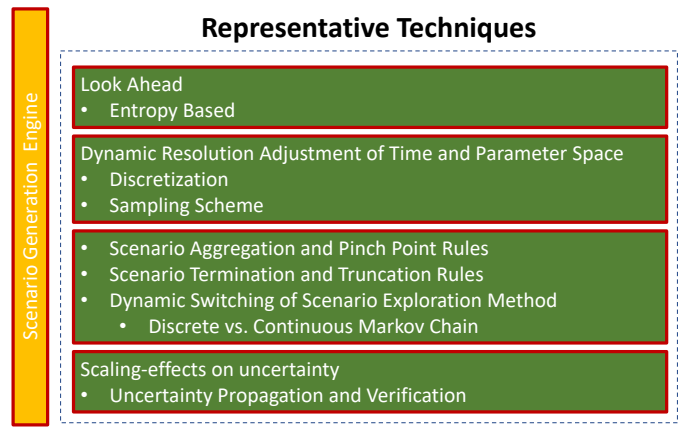

*Figure 5. Representative techniques to be integrated into the Trajectory Generation module*

## 5 PHYSICAL SIMULATION

To perform the simulation of the system at all the scale levels, various techniques can be employed to integrate the

Physical Simulation from Figure 1, such as:

## 5.1 Modular Integration Methods

Simulation of large-scale dynamic systems with modular integration methodology provides the opportunity of computational speed as well as parallel processing of individual subsystems. The concept of modular integration has been used in the analysis of large-scale stiffness analyses for decades (e.g., [23] [24] [25] [26]).

## 5.2 Graph-Theoretic Methods for Networked Dynamic Systems

Graph theory enables modeling the interconnection topology of large-scale systems and has a profound impact on the overall system in terms of its stability, controllability, observability, and performance. This method has been utilized to model a huge system and system of systems like the potable water network and power network (e.g., [27] [28] [29] [30] [31] [32]).

## 5.3 Interpolatory Model Reduction of Large-Scale Dynamical Systems

Simulations in such large-scale settings can make untenable demands on computational resources and efficient model utilization becomes necessary. Model reduction is one response to this challenge, wherein one seeks a simpler (typically lower order) model that nearly replicates the behavior of the original model. When high fidelity is achieved with a reduced-order model, it can then be used reliably as an efficient surrogate to the original, perhaps replacing it as a component in larger simulations or in allied contexts such as the development of simpler, faster controllers suitable for real-time applications (e.g., [33] [34] [35] [36] [37]).

## 5.4 Analytical Modeling by Interdependent Continuous Time Markov Chain

This method tries to compute the stochastic transition rules of a large-scale system or system of systems through historical observation and information of the system. The method provides the transition probability matrices whereby it can simulate the evolution of the system in a timely manner. The researchers have tried to mimic the stochastic behavior of large systems in different areas by continuous time Markov chain and semi-Markov models. This method utilizes data to create an analytical transition rule (e.g., [38] [39] [40]).

## 5.5 Multi-level Aggregation Methods

The method creates different levels of granularity of the system to localize and prioritize the vulnerable areas in a timely fashion. Thereafter, it can focus on the most vulnerable locations with the finer granularities [41].

## 6 CONCLUSION

In this paper, a guided simulation framework for capturing complex system behavior and guiding the simulation toward regions of high risk-significance is presented. This framework consists of an Intelligent Guidance module, a Trajectory Generation module and a System Simulation module. In this paper, several solution strategies, both qualitative and data-driven for each module are presented. In the Intelligent guidance we have presented several machine learning algorithms including Guided Simulation via Qualitative Biasing, Reinforcement Learning (RL), and Deep Reinforcement Learning (DRL). These algorithms can be selected based on the scope of the study and system complexity level.

Similar to the Intelligent Guidance, the Trajectory Generation module controls the space explosion phenomenon and interacts with the Physical Simulation using a combination of possible solution strategies. These strategies include Entropy-based Look Ahead, Exploration of Risk-Significant Regions, and Scaling-effects on uncertainty. To perform the simulation of the system at all the scale levels, various techniques will be employed to integrate the Physical Simulation modules from, such as Modular Integration Methods, Graph-Theoretic Methods for Networked Dynamic Systems, Interpolatory Model Reduction of Large-Scale Dynamical Systems, Analytical Modeling by Interdependent Continuous Time Markov Chain, and Multi-level Aggregation Methods.

The proposed framework in this study shows the interactions between the Intelligent Guidance, Trajectory Generation and System Simulation modules, that results in the capability of guided simulation for capturing complex system behavior and guided simulation toward regions of high risk-significance.